\documentclass[aps,prl,groupedaddress]{revtex4}
\usepackage{times,xspace}
\usepackage{amsbsy,amssymb,amsmath,bm}
\usepackage{amsfonts}
\usepackage{amsmath}
\usepackage{amssymb}
\usepackage[final]{graphicx}
\usepackage[figuresright]{rotating}

\begin{document}
\title{\bf The Debye-Waller factor of stabilized $\delta$-Pu}

\author{Matthias J. Graf}
\affiliation{Theoretical Division, Los Alamos National Laboratory, Los Alamos, New Mexico 87545, U.S.A. \\
(Report No.~LA-UR-06-4481 \ \ Plutonium Futures -- The Science 2006)}


\date{11 October 2007 (Available online 16 November 2006)}

\begin{abstract}
The Debye-Waller factor has been calculated for stabilized $\delta$-phase plutonium with 5\% aluminum. A quasi-harmonic Born-von K\'arman force model with temperature dependent phonon frequencies was used to calculate the mean-square thermal atomic displacement from absolute zero to 800 K. Implementation of the observed anomalous softening of the long wavelength phonons with increasing temperature cannot account for the softening of the measured thermal parameter at high temperatures nor for its rather high value at low temperatures. The implications for diffraction measurements on $\delta$-phase stabilized plutonium alloys are discussed.
\end{abstract}

\maketitle

\section{INTRODUCTION}

Since the measurement of the phonon dispersion curves along the high symmetry axes of small crystallites of $\delta$-phase $^{239}$Pu$_{0.98}$Ga$_{0.2}$ \cite{Wong2003,Wong2005}, as well as  measurements of the  phonon density of states and elastic moduli on polycrystalline  samples of $^{242}$Pu$_{0.95}$Al$_{0.05}$ \cite{McQueeney2004}, it has become possible to construct phenomenological phonon models which accurately reproduce the measured specific heat all the way up to  300 K \cite{Graf2005}. 
Unfortunately, the vibrational frequencies or density of states were determined only at a few temperatures. However, from  measurements of the elastic shear and bulk moduli it is known that the long wavelength frequencies of $\delta$-stabilized Pu exhibit an anomalously large softening, which is nearly linear from absolute zero to the melting temperature (see Fig.~1) \cite{McQueeney2004,Calder1981,Migliori2006}.

Here, I address the question of how accurately such a quasi-harmonic phonon model, with temperature dependent frequencies, can reproduce the mean-square thermal atomic displacement parameters measured by Lawson et al. \cite{Lawson1994a,Lawson1994b,Lawson2000a}. The working assumption is that all force constants, i.e, all frequencies, can be scaled by the same temperature dependent function as the observed elastic moduli or long wavelength phonons. This procedure worked very well for the calculation of the specific heat of $^{242}$Pu$_{0.95}$Al$_{0.05}$ between 2 K and 300 K \cite{Graf2005,Lashley2004}.

\section{RESULTS AND DISCUSSION}

The exponent of the Debye-Waller factor $e^{-2W({\bf q})}$ (mean-square thermal atomic displacement parameter)  is given by the average of the square of the  scalar product of the relative atomic displacement ${\bf u}$ from its average position with the neutron scattering vector ${\bf q}$, for example,  see Ref.\  \cite{Squires} for details:
$2W({\bf q}) =\langle ({\bf q \cdot u})^2 \rangle$. In the case of a monatomic cubic crystal this simplifies to the standard result for the thermal parameter 
$2W({\bf q}) = 2W_0 q^2$, where $2W_0 = \langle u_x^2 \rangle = \langle u_y^2 \rangle = \langle u_z^2 \rangle$.  The isotropic thermal parameter is given by \cite{Squires,Graf2003}
\begin{equation}
2W_0(T)= \frac{\hbar}{2 M} \int_0^\infty \frac{d\omega}{\omega} N(\omega) {\rm coth} 
\frac{\hbar \omega}{2 k_B T} \, .
\end{equation}
Here $M$ is the atomic mass, $\omega$ is the phonon frequency and the phonon density of states is normalized to satisfy $\int_0^\infty d\omega N(\omega) =1$. At temperatures much larger than the Debye temperature $\Theta_D$ the Debye-Waller temperature $\Theta_{DW}$ can easily be extracted from the  asymptotic form of  $2W_0 (T) \propto T/\Theta^2_{DW}$. For a harmonic crystal, $\Theta_{DW}$ is a constant, while Lawson et al. have found that for most light actinides $\Theta_{DW}$ acquires a temperature dependence.

\vspace{12pt}
\begin{figure}[bht]
\begin{center}
\includegraphics[width=120mm]{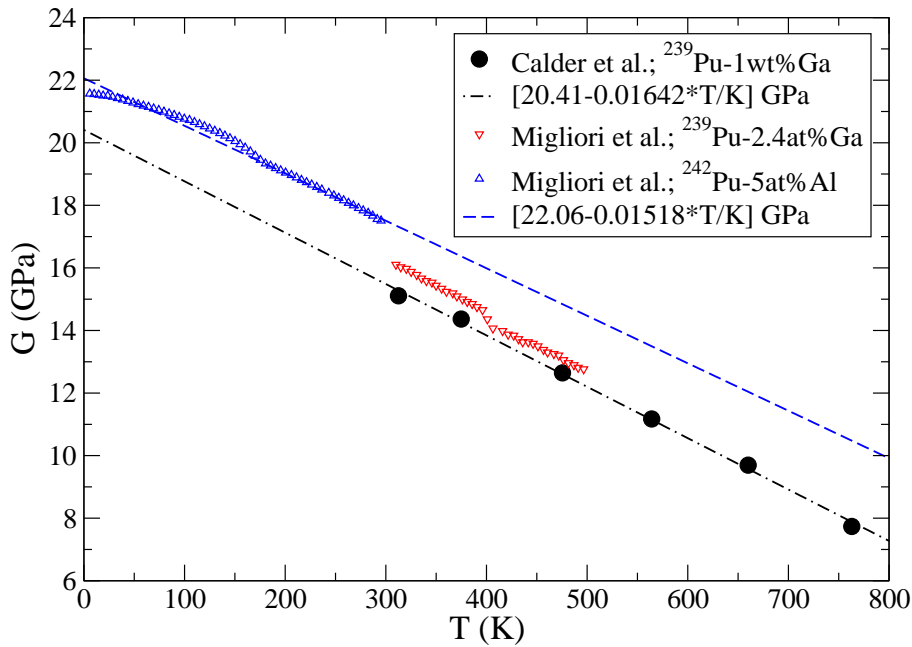}
\caption{Temperature dependence of the shear modulus $G$ of various polycrystalline $\delta$-phase Pu samples (1 wt-\% Ga $\simeq$ 3.35 at-\% Ga) \cite{McQueeney2004,Calder1981,Migliori2006}. The dashed lines are linear fits to the data of Refs.~\cite{McQueeney2004,Calder1981}.
Note a very similar temperature dependence is reported for the bulk modulus $B$.}
\end{center}
\label{fig:shear}
\end{figure}

Figure~1 shows the temperature dependence of the shear modulus $G$ of $\delta$-phase stabilized plutonium alloys. Although there is variation in the magnitude among those different alloys, their slopes are surprisingly similar. Using the temperature dependence of $G$ reported for $^{242}$Pu$_{0.95}$Al$_{0.05}$ and the 3NN (third next-neighbor) Born-von K\'arman force model discussed in Ref.~\cite{Graf2005} with parameters given in Table~\ref{table:1}, I calculated the mean-square thermal atomic displacement parameter $2W_0(T)$ as defined in Eq.~1. 
The phonon dispersions were scaled to give the following long wavelength elastic coefficients
$C_{ij}(T) = f(T) C_{ij}(300 {\rm K})$, with scaling function 
$f(T)=1.26 - 8.67\cdot 10^{-4} T/{\rm K}$, and 
$C_{11}(300 {\rm K}) = 34.0$ GPa, 
$C_{12}(300 {\rm K}) = 24.9$ GPa,  and
$C_{44}(300 {\rm K}) = 31.4$ GPa. 
Note that these single crystal moduli are consistent with the measured polycrystalline shear, $G$, and bulk, $B$,  moduli.
In Figure~2, I compare the  theoretical $2W_0(T)$, with no fit parameters, with the measured mean-square thermal atomic displacement parameters by Lawson et al. \cite{Lawson1994a}. It is worth to point out that Lawson et al.\  measured the low temperature data in a cryostat on HIPD at LANSCE (Lujan Neutron Scattering Center at Los Alamos National Laboratory) and the high temperature data in a furnace on NPD at LANSCE. For comparison I show the thermal parameters with $T$-independent force constants, i.e., $T$-independent frequencies, which were arbitrarily determined at $T=400$ K for better agreement with experiment. The calculation with the frozen-in force constants is in overall better agreement with experiment than the one with $T$-dependent force constants. This agreement is most likely fortuitous, but emphasizes the weak deviation of the experimental $2W_0$ from linearity.

\vspace{12pt}
\begin{figure}[bth]
\begin{center}
\noindent
\includegraphics[width=120mm]{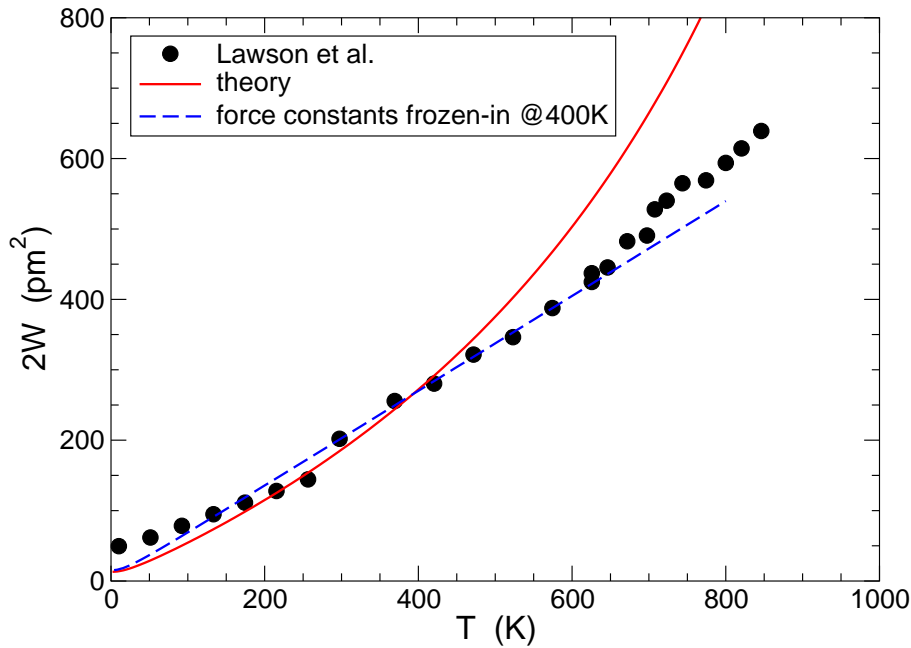}
\caption{The mean-square thermal atomic displacement parameter $2W_0(T)$ vs. temperature for Pu with 5-at.\% Al. A comparison is shown between Lawsons's data (circles) \cite{Lawson1994a} and (i) a $T$-dependent (solid line) and (ii)  $T$-independent (dashed line) 3NN Born-von K\'arman force model. For case (ii) the force constants where frozen-in by using the values of model (i) at $T=400$ K. Consequently $\Theta_{DW}$ is constant for $T > \Theta_D \approx 116$ K \cite{Graf2005}.}
\end{center}
\label{fig:2W}
\end{figure}

\begin{table}[tb]
\caption{\label{table:1}Born-von K\'arman force constants $\Phi_{ij}^{\rm nNN}$ for an {\em effective}
monatomic \textit{fcc} crystal $^{242}$Pu$_{0.95}$Al$_{0.05}$ at 300 K with
lattice parameter $a_0=4.59 {\rm \AA}$. For simplicity no correction was made to account for the 
presence of 5\% Al.}
\newcommand{\m}{\hphantom{$-$}}
\newcommand{\cc}[1]{\multicolumn{1}{c}{#1}}
\renewcommand{\tabcolsep}{2pc} 
\renewcommand{\arraystretch}{1.2} 
\begin{tabular}{@{}rrrrr}
\hline
 $\Phi_{ij}^{\rm nNN}$ (N/m) &	XX	&	ZZ	&	XY	& YZ	\\
\hline
1NN	&  8.87885	& -1.20410	& 11.92267	&		\\
2NN	& -1.29065	&  0.52554	&		&		\\
3NN	& -0.93343	&  0.03871	& -0.48034	& -0.78982	\\
\hline
\end{tabular}\\[2pt]
\end{table}

It is surprising that the realistic $T$-dependent phonon model shows the largest deviations from experiment at low and high temperatures. At high temperatures the Debye-Waller phonon moment, $\Theta_{DW}$, which corresponds to $2W_0(T)$, samples mostly the long wavelength frequencies ($\sim 1/\omega^2$) and therefore agreement should improve. Both model calculations show significant deviations from the experimental $2W_0(T)$ at low and high temperatures, while agreement is good between 180 K and 500 K.  Lawson and coworkers argued that the thermal parameter can only be determined up to an additive constant by assigning a large offset of 40-80 pm$^2$  to instrumental and systematic uncertainties in the Rietveld refinement of their neutron powder diffraction patterns. In later work  Lawson and coworkers \cite{Lawson1994b} avoided discussing the nature of this large offset all together. This offset in the thermal parameter is much larger than typically seen in diffraction measurements like fcc Ni, fcc Ce and GaAs \cite{Jeong2003,Jeong2004} when comparing experiment with simple phonon models. Although $\gamma$-Ce and $\delta$-Pu show many similarities among fcc crystals the large measured low-$T$ value for $2W_0(T)$ cannot be explained simply by scaling the hypothetical absolute zero temperature thermal parameter of $\gamma$-Ce ($2W_0(0) \approx 25$ pm$^2$) with the ratio of the atomic masses of Ce over Pu ($\sim 140/242$) \cite{Jeong2003,Jeong2004}. Here, we obtain for $\delta$-Pu the theoretical absolute zero temperature value $2W_0(0) \approx  13$ pm$^2$, which is much smaller than the experimental value of $2W_0(10 {\rm K}) \approx 49$ pm$^2$.  Therefore, the large offset observed by Lawson might be due to a combination of systematic and instrumental errors as well as  static  disorder induced broadening in the static structure function, which can lead to an overestimate of the mean-square thermal atomic displacement parameter in a Rietveld refinement. This interpretation is consistent with the more recent observation by Lawson and coworkers \cite{Lawson2000b} that there is significant contribution of diffuse scattering in neutron powder diffraction patterns of Pu$_{0.98}$Ga$_{0.02}$. Note that the low-temperature and long-wavelength phonon model successfully reproduced the low-temperature specific heat measurements \cite{Graf2005}, thus it is not clear why it should fail to explain the low-temperature thermal parameter. On the other side, the deviations seen at high temperatures could be due to anharmonicities and the breakdown of the simple scaling function used for modeling the $T$-dependent frequencies. However, the breakdown of the scaling function for the phonon dispersion is unexpected and requires further diffraction studies because it means that above $\sim 500$ K the phonons show no longer the anomalous softening as seen in the elastic moduli (see Fig.~1).

\section{CONCLUSIONS}

A 3NN Born-von K\'arman model with force constants frozen in at 400 K fortuitously reproduces the measured Debye-Waller factor over a wide temperature range, while a model with more realistic temperature-dependent phonon frequencies overestimates the mean-square thermal atomic displacement parameter above $\sim 500$ K. Both models fail to explain the large experimental low-temperature thermal parameter below $\sim 100$ K and its large asymptotic zero-temperature value ($\sim 49$ pm$^2$) due to zero-point motion. The experimental low-$T$ mean-square thermal atomic displacement parameter is nearly four times bigger than the theoretical value or the corresponding value of $\gamma$-Ce scaled by the atomic mass ratio. These discrepancies are surprising since the same temperature-dependent phonon model was very successful in describing the specific heat of $^{242}$Pu$_{0.95}$Al$_{0.05}$  between 2 K and 300 K. Further diffraction studies are needed to investigate the origin of the large deviations between theory and experiment at low and high temperatures, as well as the large unphysical offsets used in previous neutron powder diffraction measurements.

\section{ACKNOWLEDGMENTS}
I like to thank A. C. Lawson, A. Migliori, J. Mitchell  and F. Trouw for stimulating discussions.
This work was carried out under the auspices of the National Nuclear Security Administration of the U.S. Department of Energy at Los Alamos National Laboratory under Contract No. DE-AC52-06NA25396.


\begin{thebibliography}{10}

\bibitem{Wong2003} J. Wong, M. Krisch, D. L. Farber, F. Occelli, A. J. Schwartz, T.-C. Chiang, M. Wall, C. Boro, and R. Xu, Science 301 (2003) 1078.
\bibitem{Wong2005} J. Wong, M. Krisch, D. L. Farber, F. Occelli, R. Xu, T.-C. Chiang, D. Clatterbuck, A. J. Schwartz, M. Wall, and C. Boro, Phys. Rev. B 72 (2005) 64115.
\bibitem{McQueeney2004} R. J. McQueeney, A. C. Lawson, A. Migliori, T. M. Kelley, B. Fultz,  M. Ramos, B. Martinez, J. C. Lashley, and S. C. Vogel, Phys. Rev. Lett. 92 (2004) 146401.

\bibitem{Graf2005} M. J. Graf, T. Lookman, J. M. Wills, D. C. Wallace, and J. C. Lashley, Phys. Rev. B 72 (2005) 045135.

\bibitem{Calder1981}  C. A. Calder, E. C. Draney, W. W. Wilcox, J. Nucl. Mater. 97 (1981) 126.

\bibitem{Migliori2006} A. Migliori, H. Ledbetter, A. C. Lawson, A. P. Ramirez, D. A. Miller, J. B. Betts, M. Ramos, and J. C. Lashley, Phys. Rev. B 73 (2006) 52101.

\bibitem{Lawson1994a} A. C. Lawson, J. A. Goldstone, B. Cort, R. I. Sheldon, and E. M. Foltyn, in {\it Actinide Processing: Methods and Materials}; Proceedings of an international symposium held at the 123rd annual meeting of the Minerals, Metals, and Materials Society in San Francisco, California, Feb. 28 - Mar. 3, 1994; edited by B. Mishra, W. A. Averill. -- Warrendale, Pa.; (1994) 31.

\bibitem{Lawson1994b} A. C. Lawson, J. A. Goldstone, B. Cort, R. I. Sheldon, and E. M. Foltyn, J. Alloys and Comp. 213/214 (1994) 426.

\bibitem{Lawson2000a} A. C. Lawson, B. Martinez, J. A. Roberts, B. I. Bennett, and J. W. Richardson, Philos. Mag. B 80 (2000) 53.

\bibitem{Lashley2004} J. C. Lashley, J. Singleton, A. Migliori, J. B. Betts, R. A. Fisher, J. L. Smith, and R. J. McQueeney, Phys. Rev. Lett. 91 (2004) 205901.

\bibitem{Squires} G. L. Squires, {\it Introduction to the Theory of Thermal Neutron Scattering}, (Dover, Mineola, New York, 1996).

\bibitem{Graf2003} M. J. Graf, I.-K. Jeong, D. Starr, and R. H. Heffner, Phys. Rev. B 68 (2003) 064305.

\bibitem{Jeong2003} I.-K. Jeong, R. H. Heffner, M. J. Graf, and S. J. L. Billinge, Phys. Rev. B 67 (2003) 104301.

\bibitem{Jeong2004} I.-K. Jeong, T. W. Darling, M. J. Graf, Th. Proffen, R. H. Heffner, Y. Lee, T. Vogt, and J. D. Jorgensen, Phys. Rev. Lett. 92 (2004) 105702.

\bibitem{Lawson2000b} A. C. Lawson, B. Martinez, R. B. von Dreele, J. A. Roberts, R. I. Sheldon, and T. O. Brun, Philos. Mag. B 80 (2000) 1869.

\end{thebibliography}
\end{document}